\documentclass[conference]{IEEEtran}
\IEEEoverridecommandlockouts
\usepackage{cite}
\usepackage{amsmath,amssymb,amsfonts}
\usepackage{algorithmic}
\usepackage{graphicx}
\usepackage{textcomp}
\usepackage{xcolor}
\usepackage{hyperref}
\usepackage{makecell}

\hypersetup{
    colorlinks=true, 
    linkcolor=black,  
    citecolor=black, 
    filecolor=black, 
    urlcolor=black    
}

\def\BibTeX{{\rm B\kern-.05em{\sc i\kern-.025em b}\kern-.08em
    T\kern-.1667em\lower.7ex\hbox{E}\kern-.125emX}}
\begin{document}

\title{Quantum-Train-Based Distributed Multi-Agent Reinforcement Learning
\thanks{The views expressed in this article are those of the authors and do not represent the views of Wells Fargo. This article is for informational purposes only. Nothing contained in this article should be construed as investment advice. Wells Fargo makes no express or implied warranties and expressly disclaims all legal, tax, and accounting implications related to this article.\\
\IEEEauthorrefmark{1} Corresponding Author: kuan-cheng.chen17@imperial.ac.uk}
}

\author{
\IEEEauthorblockN{
    Kuan-Cheng Chen\IEEEauthorrefmark{2}\IEEEauthorrefmark{3}\IEEEauthorrefmark{1},
    Samuel Yen-Chi Chen\IEEEauthorrefmark{4},
    Chen-Yu Liu\IEEEauthorrefmark{5}\IEEEauthorrefmark{6}
    Kin K. Leung\IEEEauthorrefmark{2}\IEEEauthorrefmark{3}
}
\IEEEauthorblockA{\IEEEauthorrefmark{2}Department of Electrical and Electronic Engineering, Imperial College London, London, UK}
\IEEEauthorblockA{\IEEEauthorrefmark{3}Centre for Quantum Engineering, Science and Technology (QuEST), Imperial College London, London, UK}
\IEEEauthorblockA{\IEEEauthorrefmark{4} Wells Fargo, New York, USA}
\IEEEauthorblockA{\IEEEauthorrefmark{5}Graduate Institute of Applied Physics, National Taiwan University, Taipei, Taiwan}
\IEEEauthorblockA{\IEEEauthorrefmark{6}Hon Hai Research Institute, Taipei, Taiwan}
}

\maketitle

\begin{abstract}
In this paper, we introduce Quantum-Train-Based Distributed Multi-Agent Reinforcement Learning (Dist-QTRL), a novel approach to addressing the scalability challenges of traditional Reinforcement Learning (RL) by integrating quantum computing principles. Quantum-Train Reinforcement Learning (QTRL) leverages parameterized quantum circuits to efficiently generate neural network parameters, achieving a \(poly(\log(N))\) reduction in the dimensionality of trainable parameters while harnessing quantum entanglement for superior data representation. The framework is designed for distributed multi-agent environments, where multiple agents, modeled as Quantum Processing Units (QPUs), operate in parallel, enabling faster convergence and enhanced scalability. Additionally, the Dist-QTRL framework can be extended to high-performance computing (HPC) environments by utilizing distributed quantum training for parameter reduction in classical neural networks, followed by inference using classical CPUs or GPUs. This hybrid quantum-HPC approach allows for further optimization in real-world applications. In this paper, we provide a mathematical formulation of the Dist-QTRL framework and explore its convergence properties, supported by empirical results demonstrating performance improvements over centric QTRL models. The results highlight the potential of quantum-enhanced RL in tackling complex, high-dimensional tasks, particularly in distributed computing settings, where our framework achieves significant speedups through parallelization without compromising model accuracy. This work paves the way for scalable, quantum-enhanced RL systems in practical applications, leveraging both quantum and classical computational resources.


\end{abstract}


\section{Introduction}

Reinforcement Learning (RL) has emerged as a powerful paradigm for training agents to make sequential decisions by interacting with an environment to maximize cumulative rewards \cite{wiering2012reinforcement,sutton2018reinforcement}. Traditional RL approaches predominantly utilize classical neural networks to parameterize policies and value functions \cite{williams1992simple,mnih2016asynchronous}. However, as the complexity of tasks increases, these classical methods often face scalability issues due to the exponential growth of the parameter space required to capture intricate patterns and dependencies \cite{dulac2021challenges}.

Quantum computing (QC) offers promising solutions to overcome these limitations by harnessing quantum parallelism and entanglement, which can potentially enable more efficient processing and representation of complex data structures \cite{nielsen2010quantum}. Recent advancements in Quantum Machine Learning (QML) have demonstrated that quantum circuits can serve as powerful function approximators \cite{huang2021power, caro2022generalization}, providing theoretical speedups for specific computational tasks compared to their classical counterparts \cite{preskill2018quantum}. Variational Quantum Algorithms (VQAs) \cite{cerezo2021variational,lubasch2020variational,bharti2022noisy} form a category of hybrid machine learning approaches, where quantum parameters are optimized through classical methods. These algorithms pave the way for creating architectures like Quantum Neural Networks (QNNs), which address a wide spectrum of AI/ML challenges, including classification \cite{chen2021end, qi2023qtnvqc, mitarai2018quantum, chen2022quantumCNN, chen2024quantum, chen2024consensus,chen2024compressedmediq}, time-series forecasting \cite{chen2022quantumLSTM}, high energy physics \cite{monaco2023quantum,chen2024quantum2,chen2022quantum}, natural language processing \cite{li2023pqlm, yang2022bert, di2022dawn, stein2023applying}, and reinforcement learning \cite{chen2022variationalQRL,chen2020QRL}. These advancements also demonstrate the potential to solve real-world, large-scale, and high-dimensional problems in both supervised learning and RL.


The Quantum-Train (QT) framework employs parameterized quantum circuits to efficiently generate neural network parameters, reducing the number of trainable parameters by utilizing quantum entanglement and superposition \cite{liu2024quantum}. By leveraging the exponential size of the $n$-qubit Hilbert space, QT controls up to $2^n$ parameters using only $O(\text{poly}(n))$ rotational angles, offering significant scalability advantages. Integrated into RL tasks, this approach enhances training efficiency, particularly in distributed systems with multiple Quantum Processing Units (QPUs) \cite{liu2024qtrl,liu2024programming}. Additionally, QT eliminates common issues such as data encoding in QNNs, and its trained models are independent of quantum hardware, making it suitable for diverse real-world applications during inference stage \cite{liu2024training, lin2024quantum, liu2024federated, liu2024quantum2}.


Despite its potential benefits, integrating QT into distributed RL presents several challenges, including the synchronization of quantum-generated parameters across multiple agents, maintaining coherence in parameter updates, and ensuring convergence in the presence of quantum noise and variability. Addressing these challenges is crucial to fully realizing the potential of quantum-enhanced RL in practical applications and within the quantum-HPC ecosystem\cite{humble2021quantum,saurabh2023conceptual,chen2024quantum2}.

This paper extends the QT framework to a distributed multi-agent RL setting using multiple QPUs for parallel training and parameter synchronization. We derive the $N$-agent distributed QTRL framework, demonstrating how QT serves as a policy network parameter generator within distributed RL. Theoretical convergence analysis and proofs are provided, along with empirical results showing performance improvements. Our work advances scalable quantum-enhanced RL by leveraging quantum computational resources in a distributed environment, enabling more efficient solutions for complex, high-dimensional tasks.

\section{Reinforcement Learning Framework}

\subsection{Markov Decision Processes (MDPs)}


RL problems are commonly formulated using the framework of Markov Decision Processes (MDPs) \cite{sutton2018reinforcement}, which are defined by a tuple $\mathcal{M} = (\mathcal{S}, \mathcal{A}, \mathcal{P}, \mathcal{R}, \gamma)$. In this formalism, $\mathcal{S}$ represents a finite set of states, and $\mathcal{A}$ denotes a finite set of actions. The state transition probability function $\mathcal{P}: \mathcal{S} \times \mathcal{A} \times \mathcal{S} \rightarrow [0,1]$ defines the probability of transitioning from one state to another after taking a specific action, denoted by $\mathcal{P}(s'|s,a) = \Pr(s_{t+1} = s' \mid s_t = s, a_t = a)$. The reward function $\mathcal{R}: \mathcal{S} \times \mathcal{A} \times \mathcal{S} \rightarrow \mathbb{R}$ assigns an immediate reward for transitioning between states due to an action, and $\gamma \in [0,1)$ is the discount factor that determines the weight of future rewards. At each discrete time step $t$, the agent observes a state $s_t \in \mathcal{S}$, selects an action $a_t \in \mathcal{A}$ according to its policy $\pi(a|s)$, receives a reward $r_t = \mathcal{R}(s_t, a_t, s_{t+1})$, and transitions to a new state $s_{t+1}$ according to the probability $\mathcal{P}(s_{t+1} \mid s_t, a_t)$. The objective of the agent is to find an optimal policy $\pi^*$ that maximizes the expected cumulative discounted reward, also known as the return $\pi^* = \arg\max_{\pi} J(\pi)$, where $J(\pi) = \mathbb{E}_{\pi} \left[ \sum_{t=0}^{\infty} \gamma^{t} r_t \right]$. Here, the expectation $\mathbb{E}_{\pi}$ is taken over trajectories generated by following the policy $\pi$.

\subsection{Policy Gradient Methods}
Policy gradient methods are a class of RL algorithms that optimize the policy $\pi(a|s; \theta)$ directly by adjusting its parameters $\theta$ without the requirement to have the knowledge of transition probability $\mathcal{P}(s_{t+1} \mid s_t, a_t)$. The goal is to maximize the expected return $J(\pi)$ with respect to $\theta$. The gradient of the expected return with respect to the policy parameters is given by the policy gradient theorem \cite{sutton2018reinforcement}, $\nabla_{\theta} J(\theta) = \mathbb{E}_{\pi_{\theta}} \left[ \sum_{t=0}^{\infty} \nabla_{\theta} \log \pi(a_t \mid s_t; \theta) G_t \right]$, where $G_t = \sum_{k=t}^{\infty} \gamma^{k - t} r_k$ is the total return from time $t$ onward. This gradient can be estimated using samples from the agent's interaction with the environment, and gradient ascent methods can be employed to update the parameters $\theta$ accordingly.




\section{Quantum-Train Framework}

\subsection{Quantum Parameter Generation}

The QT framework leverages quantum computing principles to generate the parameters of classical neural networks used in RL policies. Instead of initializing and training a potentially large number of classical parameters $\theta = (\theta_1, \theta_2, \dots, \theta_k) \in \mathbb{R}^k$, QT utilizes a parameterized quantum state $|\psi(\varphi)\rangle$, where $\varphi \in \mathbb{R}^m$ represents the required $m$ parameters of a Quantum Neural Network (QNN). A quantum state with $n = \lceil \log_2 k \rceil$ qubits is prepared such that measurements in the computational basis yield probabilities corresponding to the desired parameters. Specifically, the measurement probabilities are given by, $p_i = |\langle i | \psi(\varphi) \rangle|^2$
where $i \in \{0, 1, \dots, 2^n - 1\}$ and $|i\rangle$ denotes the computational basis states. $p_i$ are the probabilities associated with each basis state upon measurement.

\subsection{Mapping Quantum Measurements to Policy Parameters}

To translate the quantum measurement probabilities into the policy parameters required by the RL algorithm, a classical mapping function $M_\beta$ is employed. This mapping is parameterized by $\beta \in \mathbb{R}^l$ and takes as input the bit-string representation $i_b$ of the basis state $|i\rangle$ and the corresponding measurement probability $p_i$, $\theta_i = M_\beta(i_b, p_i)$, where $i = 1, 2, \dots, k$. 
The mapping function $M_\beta$ can be implemented using a classical neural network or any suitable function approximator. It effectively transforms the quantum-generated probabilities into the real-valued parameters $\theta_i$ needed for the policy network.

\subsection{Policy Parameterization}

Combining the quantum parameter generation and the classical mapping, the policy parameters $\theta$ become functions of the quantum circuit parameters $\varphi$ and the mapping parameters $\beta$. We denoted it as $\theta = M_\beta(\varphi)$.
By adjusting the quantum parameters $\varphi$ and the mapping parameters $\beta$, we can indirectly modify the policy parameters $\theta$. This approach has the advantage of potentially requiring significantly fewer parameters to be trained, as the size of $\varphi$ and $\beta$ can be much smaller than $k$, that is, $m + l \ll k$,
especially when the quantum state is entangled and the mapping function is efficiently parameterized. The QT framework thus introduces a novel way of parameterizing policies in RL, leveraging quantum resources to enhance the representational capacity and potentially improve the efficiency of the learning process.

\section{Distributed Quantum-Train Reinforcement Learning}

\begin{figure*}[!t]
\centering
\includegraphics[scale=0.295]{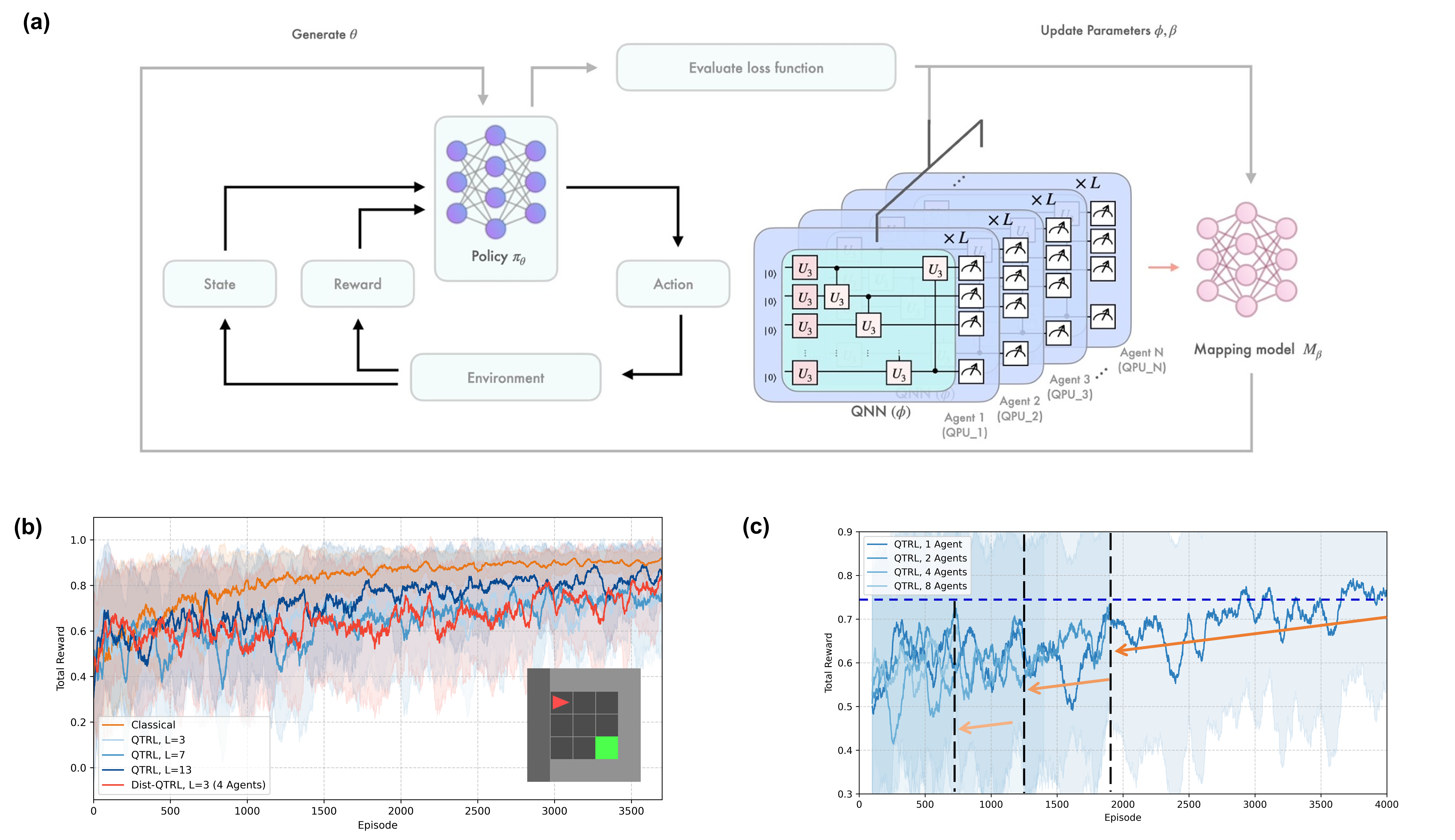}
\caption{(a) \textbf{Distributed QTRL Workflow Overview}: The workflow illustrates the interaction between the policy $\pi_\theta$, the environment, and the Graph Neural Network (GNN) module. Parameters are updated iteratively through the evaluation of the loss function, optimizing the mapping model $M_\theta$. The distributed training utilizes multiple agents, with distinct depths (L) for the GNN layers, to enhance learning efficiency. (b) \textbf{Benchmark Comparison}: This panel compares the performance of classical RL models against three centralized QTRL models with different network depths $L = [3, 7, 13]$. Results indicate that models with greater depth demonstrate better trainability, as reflected in the total reward trajectory. Notably, the distributed QTRL model with 4 agents and $L = 3$ achieves comparable performance to the centralized QTRL model with $L = 13$, showcasing the potential for distributed architectures to achieve similar efficacy with shallower networks. (c) \textbf{Speedup}: This plot highlights the speedup advantage of the distributed QTRL model. The baseline (dashed blue line) serves as a performance benchmark, indicating the number of episodes required by each model to reach the target reward of the centralized QTRL case. The distributed QTRL configurations with 2, 4, and 8 agents achieve the target performance significantly faster than centralized models, demonstrating a linear speedup due to the distributed approach.
}
\label{fig:scheme}
\end{figure*}

\subsection{Framework for Distributed Reinforcement Learning}

In a distributed RL framework, multiple agents operate concurrently, each interacting independently with its own instance of the environment. This parallelism accelerates the learning process and enhances scalability for complex tasks. In our proposed framework, we consider $N$ agents collaboratively learning an optimal policy. Each agent $i$ pulls the global shared policy $\pi_{\Theta}(a|s)$, parameterized by parameters $\Theta = \{ \beta, \varphi\}$. Agents interact with the environment independently, collecting experiences and calculating local gradients based on individual observations/experiences.
    
\subsection{Local Policy Update Mechanism}

The local policy update for each agent $i$ involves the following steps:

\begin{enumerate} \item \textbf{Experience Collection}: Agent $i$ interacts with its environment to collect a trajectory $\tau_i = {(s_t^i, a_t^i, r_t^i, s_{t+1}^i)}_{t=0}^{T_i}$, where $T_i$ denotes the length of the episode for agent $i$.

\item \textbf{Return Computation}: The agent computes the cumulative discounted return for each time step in its trajectory:
\begin{equation}
G_t^i = \sum_{k=t}^{T_i} \gamma^{k - t} r_k^i,
\end{equation}
where $\gamma \in [0,1]$ is the discount factor.

\item \textbf{Policy Gradient Estimation}: The agent estimates the policy gradient using its collected experiences:
\begin{align}
\nabla_{\Theta} J_i &= \nabla_{\Theta} J_i(M_\beta(\varphi)) \nonumber \\
&= \frac{1}{|\tau_i|} \sum_{t=0}^{T_i} \nabla_{\Theta} \log \pi_{\Theta}(a_t^i | s_t^i) \, G_t^i,
\end{align}
where $|\tau_i|$ is the number of time steps in the trajectory $\tau_i$ and $\Theta$ is the collection of all trainable parameters in QNN and the mapping model $\{ \beta, \varphi\}$.

\item \textbf{Local Gradient Computation}: The agent calculates its local gradients based on the trajectory it encounters:
\begin{equation}
    \nabla_{\Theta} J_i = \nabla_{\Theta} J_i(M_\beta(\varphi)).
\end{equation}

\end{enumerate}
Since the policy parameters $\theta$ are generated via a quantum-classical mapping $\theta = M_\beta(\varphi)$, where $\varphi$ are the quantum parameters and $\beta$ are the classical mapping parameters, each of the local agent $i$ can compute the gradient of the objective $J_{i}$ with respect to these parameters using the chain rule:

\begin{align} 
\nabla_{\Theta} J_{i}(\theta) 
&= \nabla_{\varphi, \beta} J_{i}(\theta) \nonumber \\
&= \nabla_{\varphi, \beta} J_{i}(M_\beta(\varphi)) \nonumber \\
&= \left( \frac{\partial M_\beta(\varphi) }{\partial (\varphi, \beta)} \right)^\top \nabla_{\theta} J_{i}(\theta)   
\end{align}
Each agent computes its own policy gradient $\nabla_{\theta} J_i(\theta)$ based on individual experiences. These gradients are aggregated to update the shared quantum parameters and mapping model parameters $(\varphi, \beta)$, ensuring that the quantum components of the policy benefit from the collective learning.
\subsection{Global Parameter Synchronization and Updates}
After a predefined number of episodes or iterations, agents synchronize their local gradients to converge toward a globally optimal policy. The synchronization process involves averaging the gradients across all agents:

\begin{equation} 
\Theta_{\text{new}} \leftarrow \Theta_{\text{old}} + \eta \left[ \frac{1}{N} \sum_{i=1}^N \nabla_{\Theta} J_i(M_\beta(\varphi)) \right]. 
\end{equation}
with learning rate $\eta > 0$.

Each agent then pulls the updated global shared model $\Theta_{\text{new}}$ to the local memory for next training iteration. This synchronization ensures that knowledge acquired by individual agents is shared across the network, promoting collective learning and preventing divergence due to individual biases.









In our implementation, agents perform their computations in a parallelized manner, simulating distributed quantum computing environments. Each agent interacts with its environment for one episode and computes gradients based on its experience. After all agents have completed their episodes, the gradients are aggregated, and the shared policy parameters are updated. The methods described here have the potential to be generalized to multi-QPU environments in which the QPUs are connected and can share information via classical communication networks.

\section{Result}

The performance of both classical and (Distributed) QTRL models is summarized in Tables I and II, with Fig. 1 providing further insights into their benchmarking and speedup characteristics. Table I illustrates the comparative performance of a classical policy model against various QTRL models, including both centralized and distributed architectures. While the classical policy model achieves the highest average reward of 0.9244, the proposed distributed QTRL-3 model with 4 agents performs competitively, achieving an average reward of 0.901 with significantly fewer parameters (1749) compared to the classical model's 4835. This result demonstrates the ability of distributed quantum training to reduce model complexity while maintaining high performance. Notably, the distributed QTRL-3 model achieves a comparable reward to the centralized QTRL-13, which requires a deeper QNN architecture.

\begin{table}[!t]
\centering
\caption{Comparative performance of Classical Policy and (Distributed) QT models for MiniGridEmpty-5x5-v0 environment}
\begin{tabular}{|c|c|c|c|}
\hline
\textbf{Model}                                                                                             & \textbf{Topology}                                                                                      & \textbf{\begin{tabular}[c]{@{}c@{}}Average \\ Reward\end{tabular}} & \textbf{Para. Size} \\ \hline
\begin{tabular}[c]{@{}c@{}}Classical \\ Policy Model\end{tabular}                                          & (147-32, 32-3)                                                                                         & 0.9244                                                             & 4835                \\ \hline
Centric QTRL-3                                                                                             & \begin{tabular}[c]{@{}c@{}}QNN: (U3 block)*3 \\ + Mapping Model: \\ (11-10, 10-10, 10-1)\end{tabular}  & 0.6453                                                             & 1749                \\ \hline
Centric QTRL-7                                                                                             & \begin{tabular}[c]{@{}c@{}}QNN: (U3 block)*7 \\ + Mapping Model: \\ (11-10, 10-10, 10-1)\end{tabular}  & 0.8164                                                             & 2061                \\ \hline
Centric QTRL-13                                                                                            & \begin{tabular}[c]{@{}c@{}}QNN: (U3 block)*13 \\ + Mapping Model: \\ (11-10, 10-10, 10-1)\end{tabular} & 0.8632                                                             & 2529                \\ \hline
\textbf{\begin{tabular}[c]{@{}c@{}}Distributed QTRL-3 \\ (4 Agents) \\  {[}Proposed Model{]}\end{tabular}} & \begin{tabular}[c]{@{}c@{}}QNN: (U3 block)*3 \\ + Mapping Model: \\ (11-10, 10-10, 10-1)\end{tabular}  & 0.901                                                              & 1749                \\ \hline
\end{tabular}
\label{tab:comparative_qt_models}
\end{table}

Table II highlights the speedup achieved by distributed QTRL models compared to their centralized counterparts. As seen, the distributed QTRL-3 model with 2 agents provides a speedup factor of 2.06, while configurations with 4 and 8 agents yield speedups of 3.33 and 5.33, respectively. These results indicate a near-linear scaling of computational efficiency as the number of agents increases, confirming the efficacy of the distributed QTRL approach. Fig. 1(c) further validates this finding by demonstrating that the distributed QTRL models consistently reach the target reward in significantly fewer episodes compared to their centralized counterparts, thereby underscoring the advantage of distributed quantum training in RL tasks.

\begin{table}[!t]
\centering
\caption{Speed-up comparison of centric and distributed QTRL models}
\begin{tabular}{|c|c|c|}
\hline
\textbf{Model}                                                                                             & \textbf{Topology}                                                                                      & \textbf{\begin{tabular}[c]{@{}c@{}}Speed-up \end{tabular}} \\ \hline
\textbf{\begin{tabular}[c]{@{}c@{}}Centric QTRL-3 \\ (1 Agent) - Proposed Model\end{tabular}}               & \begin{tabular}[c]{@{}c@{}}QNN: (U3 block)*3 \\ + Mapping Model: \\ (11-10, 10-10, 10-1)\end{tabular}  & -                                                                                      \\ \hline
\textbf{\begin{tabular}[c]{@{}c@{}}Distributed QTRL-3 \\ (2 Agents) - Proposed Model\end{tabular}}          & \begin{tabular}[c]{@{}c@{}}2*[QNN: (U3 block)*3] \\ + Mapping Model: \\ (11-10, 10-10, 10-1)\end{tabular} & x2.06                                                                                 \\ \hline
\textbf{\begin{tabular}[c]{@{}c@{}}Distributed QTRL-3 \\ (4 Agents) - Proposed Model\end{tabular}}          & \begin{tabular}[c]{@{}c@{}}4*[QNN: (U3 block)*3] \\ + Mapping Model: \\ (11-10, 10-10, 10-1)\end{tabular} & x3.33                                                                                 \\ \hline
\textbf{\begin{tabular}[c]{@{}c@{}}Distributed QTRL-3 \\ (8 Agents) - Proposed Model\end{tabular}}          & \begin{tabular}[c]{@{}c@{}}8*[QNN: (U3 block)*3] \\ + Mapping Model: \\ (11-10, 10-10, 10-1)\end{tabular} & x5.33                                                                                 \\ \hline
\end{tabular}
\label{tab:speedup_comparison}
\end{table}

\section{Conclusion}
This study presents the Dist-QTRL framework, which effectively integrates distributed quantum computing with reinforcement learning to address scalability and performance challenges in high-dimensional environments. By leveraging parameterized quantum circuits, the QTRL framework enables efficient policy generation, reduces trainable parameters compared to the classical RL model, and achieves significant speedups in training through parallelization across multiple QPUs (compared to centric QTRL with one QPU). Theoretical analyses and empirical results demonstrate the framework’s convergence and highlight its superiority over classical approaches in terms of both accuracy and scalability. This work contributes to advancing scalable quantum-enhanced RL systems, laying a foundation for more efficient and robust applications in complex, real-world scenarios. Future work will focus on quantum circuit partitioning for large-scale distributed quantum computing through the distributed circuit ansatz\cite{liu2024quantum,liu2024quantum3,burt2024generalised}, as well as exploring differentiable quantum architecture search in asynchronous quantum reinforcement learning\cite{chen2024differentiable} to further enhance scalability and performance.

\section*{Acknowledgment}
This research was funded by the EPSRC Distributed Quantum Computing and Applications project (grant number EP/W032643/1).

\bibliographystyle{ieeetr}
\bibliography{references,bib/rl,bib/qc,bib/qml_examples}

\end{document}